\documentclass[twoside,a4paper,11pt,reqno]{amsart}
\usepackage{amsmath,mathrsfs}
\usepackage{tabularx}
\numberwithin{equation}{section}

\def\cal{\mathcal}
\newtheorem{theorem}{Theorem}
\newtheorem{lemma}{Lemma}
\newtheorem{definition}{Definition}

\newtheorem{remark}{Remark}
\numberwithin{equation}{section}

\begin{document}
\title[Global solution to the
Cauchy problem on UFM] {Global solution to the Cauchy problem on a
universe fireworks model}

\author{Zhenglu Jiang}
\address{Department of Mathematics, Zhongshan University,
Guangzhou 510275, China} \email{mcsjzl@mail.sysu.edu.cn}
\thanks{This work  was supported by NSFC 10271121 and
the Scientific Research Foundation for the Returned Overseas Chinese
Scholars, the Ministry of Education of China, and sponsored by joint
grants of NSFC 10511120278/10611120371 and RFBR 04-02-39026.}

\author{Hongjiong Tian}
\address{Department of Mathematics,  Shanghai Teachers' University,
Shanghai 200234,  China} \email{hongjiongtian@263.net}


\date{\today.}



\begin{abstract}
We prove  existence and uniqueness of the global solution to the
Cauchy problem on a universe fireworks model with finite total mass
at the initial state when the ratio of the mass surviving the
explosion, the probability of the explosion of fragments and the
probability function of the velocity change of a surviving particle
satisfy the corresponding physical conditions. Although the
nonrelativistic Boltzmann-like equation modeling the universe
fireworks is mathematically easy, this paper leads rather
theoretically to an understanding of how to construct contractive
mappings in a Banach space for the proof of the existence and
uniqueness by means of methods taken from the famous work by DiPerna
\& Lions about the Boltzmann equation. We also show both the
regularity and the time-asymptotic behavior of solution to the
Cauchy problem.
\end{abstract}

\maketitle

\section{Introduction}
\label{intro} We are concerned with existence and uniqueness of the
global solution to the Cauchy problem on a universe fireworks model
(hereafter, UFM) without gravity described by the following
nonrelativistic
 Boltzmann-like equation \cite{wl}
\begin{equation}
\frac{\partial f}{\partial t}+\xi \frac{\partial f}{\partial x}=
\eta(t, x, \xi)\int_{{\bf R}^3}\Gamma(t, x, \xi_1)P(-\xi\cdot\xi_1)f_1d\xi_1
-\Gamma(t, x, \xi)f
\label{(2.1.1)}
\end{equation}
where $f\equiv f(t, x, \xi)$ is the total
mass distribution with respect to time
$t$,  space $x$ and velocity $\xi$ during the explosion of a huge cloud
of matter; $f_1\equiv f(t, x, \xi_1)$; $\eta\equiv
\eta(t, x, \xi)$ is the mass ratio which is a measure of the mass
surviving the explosion; $\Gamma\equiv
\Gamma(t, x, \xi)$ is the explosion probability per unit time for a fragment;
$P\equiv P(-\xi\cdot\xi_1)$
 is the probability density so that $P(-\xi\cdot\xi_1)d\xi_1$
 is the probability that a surviving particle will change its
 velocity  $\xi$ to $\xi_1\in d\xi_1$ during explosion.

Assume that $\Gamma,  \eta$ and $P$ satisfy the following physical
conditions
\begin{equation}
0\leq\eta\leq 1,  0\leq\Gamma\leq 1, 0{\leq} P,
\label{(2.1.2)}
\end{equation}
\begin{equation}
\int_{{\bf R}^3} P(-\xi\cdot\xi_1)d\xi_1=1 \hbox{  for  }
\forall \xi\in {{\bf R}^3},
\label{(2.1.3)}
\end{equation}
and that the initial total mass distribution $f_0\equiv f_0(x, \xi)$ satisfies
\begin{equation}
 0\leq f_0(x, \xi)\in L^1({{\bf R}^3}
\times{{\bf R}^3}),
\label{(2.1.4)}
\end{equation}
i.e., having finite total mass at the initial state.

In this paper we shall prove existence and uniqueness of the global
solution to the Cauchy problem on the nonrelativistic Boltzmann-like
equation~(\ref{(2.1.1)}) of the distribution $f$ with finite total
mass at the initial state when the ratio $\eta,$ the probability
$\Gamma $  and the probability density $P$ satisfy the corresponding
physical conditions (\ref{(2.1.2)}) and (\ref{(2.1.3)}). We shall
also show both the regularity and the time-asymptotic behavior of
solution to this problem.

The nonrelativistic Boltzmann-like equation is mathematically almost
trivial since it is a linear transport equation with an integrable
kernel and bounded multiplicative coefficients. However, this paper
leads rather theoretically to an understanding of how to construct
contractive mappings in a Banach space by means of methods taken
from the famous work by DiPerna \& Lions \cite{dl} about the
Boltzmann equation.

The fireworks model is one of the metagalaxy models which are used
to give a physical explanation of the expansion. Alfv\'{e}n
\cite{ah} suggested the first fireworks model for the evolution of
the universe
 as an alternative cosmology to big bang and described
a possible mechanism for this model in an original work in 1983, but
many details remain unexplained because of the extremely complicated
physics involved \cite{wl}. Some assumptions on the mechanism of the
explosion, such as  that matter and antimatter are mixed, are
invalid in a metagalaxy model given by Alfv\'{e}n \& Klein \cite{ak}
for the highest redshift observed. Also, the highest redshift
observed cannot be accounted for even if it is assumed that there is
an upper limit to the velocities that could be reached by one
explosion of about $0.7c$ ($c$ is the light velocity) \cite{lbc}. A
special relativistic model was investigated by Laurent \& Carlqvist
\cite{lc} but the evolution of the distributions in configuration
space was not extended to general relativity by use of their
approach. Due to these, Widlund \cite{wl} gave a fireworks model
(defined by equation~(\ref{(2.1.1)}))  without prescribing the
details of the explosion mechanism and developed the model using a
fully general relativistic treatment of the fireworks. The
nonrelativistic fireworks model, which is given by
equation~(\ref{(2.1.1)}), is a special case of the nonrelativistic
limit to the general relativistic model considered by Widlund
\cite{wl}. Widlund has explained the observed redshifts only from
the start of the study of an approximate equation of
equation~(\ref{(2.1.1)}) because it is a differential equation and
has an exact solution for
 many weak explosions. Our results mentioned above are not only a kind of
 reasoning to adopt the research method of replacing equation~(\ref{(2.1.1)})
 by an approximate equation  as Widlund has done, but additionally
 make it possible to do some
  research work about the redshifts directly from the start of
   the solution to equation~(\ref{(2.1.1)}).

\section{Existence and Uniqueness}
One of our results mentioned in the previous section can be
described as follows.
\begin{theorem}
Equation~(\ref{(2.1.1)}) with physical conditions (\ref{(2.1.2)})
and (\ref{(2.1.3)}) has a unique nonnegative global distributional
solution $f(t,x, \xi)$ through a total mass distribution $f_0(x,
\xi)$ satisfying a finite total mass condition (\ref{(2.1.4)}) at
the initial state, and this solution $f(t,x, \xi)$ belongs to the
Banach space $ L^\infty((0,+\infty);L^1({\bf R}^3\times{\bf R}^3)).$
\label{th01}
\end{theorem}

Theorem~\ref{th01} shows existence and uniqueness of solution to the
Cauchy problem on UFM without gravity no matter how long the
fireworks era lasts.

The proof of Theorem~\ref{th01} is as follows. We first define a
mild solution to equation~(\ref{(2.1.1)}) as DiPerna \& Lions did in
\cite{dl}. Then,  we establish a Banach space and a mapping of this
space into itself which uniformly decreases distances,  and shows
that this mapping has a unique fixed point which is a unique global
mild solution to equation~(\ref{(2.1.1)}). This is an application of
 the well-known Banach fixed point theorem. Finally, according to
 the relations between mild solution
 and distributional solution (e.g., \cite{dl}, \cite{j98a}),
 we know that this mild solution is also
 a global distributional solution.

\begin{definition}
Let $f=f(t, x, \xi)$ be a nonnegative function which belongs to
$L_{loc}^1((0, T)\times{{\bf R}^3}\times{{\bf R}^3}), $ and assume
that for almost all $(x, \xi)\in{{\bf R}^3}\times{{\bf R}^3}, $
\begin{equation}
Q^\#(t, x, \xi)\in L^1(0, T_1)\hbox{  } (\forall T_1\in(0, T))
\label{(2.2.0)}
\end{equation}
and
\begin{equation}
f^\#(t, x, \xi)-f^\#(s, x, \xi)=\int^t_sQ^\#(\sigma, x, \xi)d\sigma\hbox{  }
(\forall 0{\leq}s<t{\leq}T),
\label{(2.2.1)}
\end{equation}
where $h^\#$ denotes, for any measurable function $h$ on $(0, +\infty)
\times{{\bf R}^3}\times{{\bf R}^3}, $
the following restriction to characteristics:
$$h^\#(t, x, \xi)=h(t, x+t\xi, \xi).$$
If $0<T<+\infty, $ $f$ is called a local-in-time mild solution to
the equation
\begin{equation}
 \frac{\partial f}{\partial t}+\xi \frac{\partial f}{\partial x}=Q
\label{(2.2.2)}
\end{equation}
in the time interval $[0, T].$ If $T=+\infty, $ $f$ is called a global
 mild solution to equation~(\ref{(2.2.2)}).
\label{def01}
\end{definition}

By Definition~\ref{def01},  we have the following result:

\begin{lemma}
Suppose that $f, h\in L_{loc}^1({\bf R}\times{{\bf R}^3}\times{{\bf
R}^3}), $ and that for almost all $x, \xi\in{{\bf R}^3}, $ $f^\#$ is
absolutely continuous with respect to $t, $ $h^\#\in L_{loc}^1({\bf
R}).$ Then $f$ is a distributional solution to
equation~(\ref{(2.2.2)}) if and only if $f$ is a mild solution to
equation~(\ref{(2.2.2)}). \label{lemma1}
\end{lemma}

This result is similar to that given by DiPerna \& Lions \cite{dl}
for the Boltzmann equation. To show Theorem~\ref{th01}, we also have
to define a mapping as follows.

\begin{definition}
A mapping $J$ is defined as follows: for any $
f\in {\cal F}$ and $t\in [0, T], $
$$
J(f)(t, x, \xi)=f_0(x-t\xi, \xi)-\int^t_0\Gamma(\sigma, x+(\sigma-t)\xi, \xi)
f(\sigma, x+(\sigma-t)\xi, \xi)d\sigma $$
\begin{equation}
+\int^t_0\int_{{\bf R}^3}\eta(\sigma, x+(\sigma-t)\xi, \xi)\Gamma(\sigma, x+(\sigma
-t)\xi, \xi_1)P(-\xi\cdot\xi_1)f(\sigma, x+(\sigma-t)\xi, \xi_1)d\xi_1d\sigma
  \label{(2.2.3)}
\end{equation}
 where ${\cal F}=\left\{f\equiv f(t, x, \xi):
f\in L^\infty((0, T);L^1({{\bf R}^3}\times{{\bf R}^3}))
 \hbox{ for all  }  T\in [0, +\infty)\right\}.$
\label{def02}
\end{definition}

Let the Lebesgue measure be used for all the integrals in this paper.
Since the Lebesgue measure defined in ${\bf R}\times{\bf R}^3\times{\bf R}^3$
 is $\sigma$-finite,  by use of the Fubini-Tonelli Theorem
and the physical condition (\ref{(2.1.3)}),
we know
that $f, f^\#,  \int_{{\bf R}^3}P(-\xi\cdot\xi_1)f_1d\xi_1,
\left[\int_{{\bf R}^3}P(-\xi\cdot\xi_1)f_1d\xi_1\right]^\#
\in L_{loc}^1(0, +\infty)$ for almost every $(x, \xi)\in
{\bf R}^3\times{\bf R}^3, $ i.e.,  the product $P(-\xi\cdot\xi_1)f(\cdots)$
 in equation (\ref{(2.2.3)}) is integrable with respect to $\xi_1.$
 Thus Definition~\ref{def02} is valid.

Take $T_0$ as a definite time the fireworks era lasts. Then, if $0\leq T_0
<\frac{1}{2},$   equation~(\ref{(2.1.1)}) has a local-in-time mild solution in the time interval
$[0, T_0], $ or say,  the total mass distribution of the universe fireworks is
uniquely determined by the initial total mass data during the  fireworks
 era. Indeed,  by Definition~\ref{def02} and with the help of the physical conditions
(\ref{(2.1.2)}) and (\ref{(2.1.3)}),  we can
 easily prove that $J(f)\in {\cal F}$ for all $ f\in {\cal F}, $
 and that the following
 inequality holds:
\begin{equation}
\mathop{\max}\limits_{0\leq t \leq T_0}\|J(f)-J(h)\|_{L^1({{\bf R}^3}\times{{\bf R}^3})}
 \leq 2T_0\mathop{\max}\limits_{0\leq t \leq T_0}\|f-h\|_{L^1(
 {{\bf R}^3}\times{{\bf R}^3})}
\label{(2.2.4)}
\end{equation}
 for all $ f, h \in {\cal F}.$ The inequality (\ref{(2.2.4)}) shows that $J$ is a contractive mapping from
 ${\cal F}$ into itself with the same norm as in
 $C((0, T_0);L^1({\bf R}^3\times{\bf R}^3)).$
  Therefore there exists a unique
  element $h_1\in {\cal F}$ such that $J(h_1)=h_1$ for a.e. $(t, x, \xi)\in
  (0, T_0) \times{{\bf R}^3}\times{{\bf R}^3}.$
  In order to prove that $h_1$ is a local-in-time mild
  solution to equation~(\ref{(2.1.1)}),  it is enough to show that $h_1$ is nonnegative for almost
  every $(t, x, \xi)\in (0, T_0)\times{{\bf R}^3}\times{{\bf R}^3}.$

 Let us take another mapping $J_1$ as follows: $J_1(f)=\max(0, J(f))$ for all
$ f \in {\cal F}^+=\{f:f\in {\cal F} \hbox{ and }
f\mathop{\geq}\limits^{a.e.}0\}.$ Obviously,
 ${\cal F}^+$ is a subset of ${\cal F}.$ Similarly, we can easily show that the
 mapping $J_1$ maps ${\cal F}^+$ into itself and is uniformly contractive
  with the same norm as mentioned above.
  Then there exists a unique element $f_1\in {\cal F}^+$
  such that $J_1(f_1)=f_1$ for almost every $(t, x, \xi)\in[0, T_0]\times
  {{\bf R}^3}\times{{\bf R}^3}.$ Thus,  if $f_1=J(f_1)$ for almost every $(t, x, \xi)\in [0, T_0]
  \times{{\bf R}^3}\times{{\bf R}^3}, $ $f_1$ is a local-in-time mild solution to equation~(\ref{(2.1.1)}) through
  $f_0$ in the time interval $[0, T_0], $ and $f_1\mathop{=}\limits^{a.e.}h_1.$
   We will below show that $J(f_1)\mathop{=}\limits^{a.e.}f_1, $ or equivalently,
   $J(f_1)^\#\mathop{=}\limits^{a.e.}f_1^\#.$  In fact,  by equation (\ref{(2.2.3)}),  we know that
   $J(f_1)^\#$ is absolutely continuous with respect to $t\in [0, T_0]$ for
   almost every $(x, \xi)\in {{\bf R}^3}\times{{\bf R}^3}.$ We may assume without loss of
    generality  that $J(f_1)^\#(t, x, \xi)$ is continuous for all $(t, x, \xi).$
To prove that $J(f_1)^\#=f_1^\#, $ it suffices to prove that $J(f_1)^\#
\geq 0.$ If $J(f_1)^\#(t_0, x_0, \xi_0)<0$ for some point $(t_0, x_0, \xi_0)
\in [0, T_0]\times{{\bf R}^3}\times{{\bf R}^3}, $ we can find some value $t_1 \in [0, T_0]$
such that $J(f_1)^\#(t, x_0, \xi_0)<0$ for all $t\in[t_1, t_0], $ so that $
f_1^\#(t, x_0, \xi_0)=0$ for all $t\in [t_1, t_0].$ By equation (\ref{(2.2.3)}),  we can know that
$$0>J(f_1)^\#(t, x_0, \xi_0)\geq J(f_1)^\#(t_1, x_0, \xi_0)~~(t_0>t>t_1).$$
Repeating the above analysis,  we can conclude that
$J(f_1)^\#(0, x_0, \xi_0)<0, $ i.e.,  $f_0(x_0, \xi_0)<0, $ which is a
contradiction. Therefore $h_1=f_1=J(f_1)\geq 0$ for almost every $
(t, x, \xi)\in [0, T_0]\times{{\bf R}^3}\times{{\bf R}^3}.$

How about $T_0\geq\frac{1}{2}?$ It suffices to consider the
$T_0=+\infty$ case,  i.e.,  to prove Theorem~\ref{th01}. In order to
do this,  we have to establish
 a subset ${\cal B}$ of ${\cal F}, $ such that $J$ is a contractive
 mapping from ${\cal B}$
 into itself with some norm in the time interval
 $[0, +\infty).$ Now,  let us take ${\cal F}_a$ as follows:
 $${\cal F}_a=\left\{ f: \|f\|_a\equiv \mathop{\sup}\limits_{0\leq t<+\infty}
 \left\{e^{-at}\int\int_{{{\bf R}^3}\times{{\bf R}^3}}|f(t, x, \xi)|dxd\xi
 \right\}\leq A,
 f\in {\cal F}\right\}$$
 where $a>0, A=
 \frac{a}{2}\|f_0\|_{L^1({{\bf R}^3}\times{{\bf R}^3})}.$
 Then $({\cal F}_a, \|\cdot\|_a)$ is a complete Banach space. By Definition~\ref{def02} and
 with the help of the physical conditions (\ref{(2.1.2)}) and (\ref{(2.1.3)}),  it is also easy to see that
 \begin{equation}
 \|J(f)\|_a\leq A,
\label{(2.2.5)}
\end{equation}
\begin{equation}
\|J(f)-J(h)\|_a\leq \frac{2}{a}\|f-h\|_a.
\label{(2.2.6)}
\end{equation}
 for all $f, h \in {\cal F}_a.$
 We know from the two inequalities (\ref{(2.2.5)}) and (\ref{(2.2.6)})
that $J$ is a contractive mapping of
  ${\cal F}_a$ into itself with the norm $\|\cdot\|_a$ in
  the time interval $[0, +\infty)$
  for any given $a>2.$ Thus,  choosing a real $a_0$ such that $a_0>2$ and
  taking ${\cal B}={\cal F}_{a_0}, $ we can have a unique element $h_0\in {\cal B}$ such that
  $J(h_0)=h_0$ in $[0, +\infty)\times{\bf R}^3\times{\bf R}^3.$
  As stated above about the fixed point nonnegativity in the
 $0\leq T_0<\frac{1}{2}$ case,  we can easily show that $h_0$ is nonnegative for
 almost every $(t, x, \xi)\in [0, +\infty)\times{{\bf R}^3}\times{{\bf R}^3}.$
Therefore equation (\ref{(2.1.1)}) with the physical conditions
(\ref{(2.1.2)}) and (\ref{(2.1.3)}) has a global mild solution
through the total mass distribution $f_0$ satisfying the finite
total mass condition (\ref{(2.1.4)}) at the initial state. Finally,
by Lemma~\ref{lemma1},  Theorem~\ref{th01} holds true.

We can also construct another contractive mapping to prove
Theorem~\ref{th01}. Let us write
$F^\#(t)=\int_0^t\Gamma^\#(\sigma,x,\xi)d\sigma. $ Then, using the
same device as given by DiPerna \& Lions \cite{dl} for the Boltzmann
equation, we have
\begin{lemma}
 $f=f(t,x,\xi)$ is a
mild solution to equation~(\ref{(2.1.1)}) if and only if
$f=f(t,x,\xi)$ satisfies
$$f^\#(t,x,\xi)-f^\#(s,x,\xi)e^{-[F^\#(t)-F^\#(s)]} $$ $$
=\int_s^t \int_{{\bf R}^3}\eta^\#(\sigma, x, \xi)\Gamma^\#(\sigma, x, \xi_1)P(-\xi\cdot\xi_1)f^\#(\sigma,x, \xi_1)
e^{-[F^\#(t)-F^\#(\sigma)]} d\xi_1 d\sigma $$
for almost all $(x,\xi) \in {\bf R}^3\times{\bf R}^3$ and all $0<s<t<+\infty.$
\label{lemma2}
\end{lemma}
\begin{definition}
A mapping $J^+$ is defined as follows:  $\forall
f\in {\cal F}^+$ and $\forall  t\in [0, T], $
$$ J^+(f)(t, x, \xi)=e^{-\int_0^t\Gamma(\sigma,x+(\sigma-t)\xi,\xi)d\sigma}[f_0(x-t\xi, \xi) $$
\begin{equation}
+\int^t_0Q^+(\sigma,x+(\sigma-t)\xi, \xi_1)
e^{-\int_0^\sigma\Gamma(\tau,x+(\tau-t)\xi,\xi)d\tau}d\sigma],
  \label{(2.2.7)}
\end{equation}
 where $${\cal F}^+=\left\{f\equiv f(t, x, \xi): f\geq 0 \hbox{ and }
f\in L^\infty((0, T);L^1({{\bf R}^3}\times{{\bf R}^3}))
 \hbox{ for all }  T\in [0, +\infty) \right\},$$
$$ Q^+(t,x, \xi)
=\int_{{\bf R}^3}\eta(t, x, \xi)\Gamma(t, x, \xi_1)P(-\xi\cdot\xi_1)f(t,x, \xi_1)d\xi_1.$$
\label{def03}
\end{definition}

Using Lemma \ref{lemma2} and Definition \ref{def03}, we can give a
different and brief proof of Theorem~\ref{th01}. Let us take ${\cal
F}_a^+$ as follows:
 $${\cal F}_a^+=\left\{ f: \|f\|_a\equiv \mathop{\sup}\limits_{0\leq t<+\infty}
 \left\{e^{-at}\int\int_{{{\bf R}^3}\times{{\bf R}^3}}|f(t, x, \xi)|dxd\xi
 \right\}\leq A,
 f\in {\cal F}^+\right\}$$
 where $a>0, A=
 a\|f_0\|_{L^1({{\bf R}^3}\times{{\bf R}^3})}.$
 Then $({\cal F}_a^+, \|\cdot\|_a)$ is a complete metric space
with the distance $dist(\cdot,\cdot)=\|\cdot-\cdot\|_a.$
 Under the physical conditions (\ref{(2.1.2)}) and (\ref{(2.1.3)}),
it can be easily seen from Definition~\ref{def03} that
 \begin{equation}
 \|J^+(f)\|_a\leq A,
\label{(2.2.8)}
\end{equation}
\begin{equation}
\|J^+(f)-J^+(h)\|_a\leq \frac{1}{a}\|f-h\|_a.
\label{(2.2.9)}
\end{equation}
 for all $f, h \in {\cal F}_a^+.$ By (\ref{(2.2.8)}) and (\ref{(2.2.9)}),
it follows that $J^+$ is a contractive mapping from
  ${\cal F}_a^+$ into itself with the distance $\|\cdot-\cdot\|_a$ in
  the time interval $[0, +\infty)$
  for any given $a>1.$ This implies that if we put $a>1,$ then
there exists a unique element $g_0\in {\cal F}_a^+$ such that
$J^+(g_0)=g_0.$  Hence, by Lemma~\ref{lemma2}, Theorem~\ref{th01}
holds true.

Similarly,
 we can further show the regularity of solution to the Cauchy
problem, that is,
\begin{theorem}
Let $k$ be any fixed natural and $f$ a unique nonnegative global
distributional solution under assumptions in Theorem~\ref{th01}.
Support that $\eta$ and $\Gamma$ are in $C^k([0,+\infty)\times{\bf
R}^3\times{\bf R}^3)$ and that $P\in C^k({\bf R}).$ If $f_0\in
C^k({\bf R}^3\times{\bf R}^3),$ then $f\in C^k([0,+\infty)\times{\bf
R}^3\times{\bf R}^3).$ \label{th02}
\end{theorem}

In fact, under all the assumptions in Theorem~\ref{th02}, we can
find that $J^+$ defined by (\ref{(2.2.7)}) is still a contractive
mapping from ${\cal F}_a^+\cap C^k([0,+\infty)\times{\bf
R}^3\times{\bf R}^3)$ into itself for any fixed $a>1,$ and then
easily know that Theorem~\ref{th02} holds.

\begin{remark} In the relativistic case,  the total rest mass
distribution of UFM without gravity is uniquely determined by the
initial total rest mass data as in the nonrelativistic case. Since
the velocity $\xi$ in the relativistic case is not greater than the
light velocity $c$ (for convenience,  write $c=1$),  we
 have to choose a transformation: $\xi=\frac{p}{\sqrt{1+p^2}}$ where $p$ is
a variable of momentum for the relativistic UFM, and regard the
 total rest mass distribution as that of the three variables of time $t,$ space $x$ and momentum $p.$
Thus we can easily show the results mentioned above as done in the
nonrelativistic case,
  although the form of the transport operator $\frac{\partial}{\partial t}
  +\frac{p}{\sqrt{1+p^2}}\frac{\partial}{\partial x}$ of the relativistic
  Boltzmann-like equation \cite{wl} is different from that of
the nonrelativistic Boltzmann-like equation (\ref{(2.1.1)}). It is
here worthwhile to mention that many results about the Cauchy
problems relative to the relativistic Boltzmann or Enskog equation
have been given in a number of papers (e.g., \cite{g06}, \cite{j99},
\cite{j07a}, \cite{j07b}).
\end{remark}

\section{Asymptotic Behavior}
In this section we shall study the time-asymptotic behavior of the
solution to the Cauchy problem on UFM. We can find that this
solution is time-asymptotically convergent to the free motion in
$L^1$-norm under some suitable assumptions.

To do so, we first have to show the following lemma:
\begin{lemma}
Let $\delta$ be a positive constant less than $1.$  In addition to
all the assumptions in Theorem~\ref{th01}, if $\eta$ and $P$ satisfy
the following inequality:
\begin{equation}
\int_{{\bf R}^3}\eta(t,x,\xi)P(-\xi\xi_1)d\xi\leq\delta
 \label{con04}
\end{equation}
for all the three variables $(t,x,\xi_1),$ then the nonnegative
solution $f$ to the Boltzmann-like equation (\ref{(2.1.1)})
satisfies
\begin{equation}
\frac{d}{dt}\iint_{{\bf R}^3\times{\bf R}^3}f(t,x,\xi)dxd\xi
+(1-\delta)\iint_{{\bf R}^3\times{\bf
R}^3}\Gamma(t,x,\xi)f(t,x,\xi)dxd\xi\leq 0.
 \label{ineqfun01}
\end{equation}
Furthermore, we have
\begin{eqnarray}
\iint_{{\bf R}^3\times{\bf R}^3}f(s,x,\xi)dxd\xi \hspace{8cm}\nonumber \\
\leq(\delta-1)\int_0^s\iint_{{\bf R}^3\times{\bf
R}^3}\Gamma(t,x,\xi)f(t,x,\xi)dtdxd\xi+ \iint_{{\bf R}^3\times{\bf
R}^3}f_0(x,\xi)dxd\xi
 \label{ineqfun02}
\end{eqnarray}
for any $s\in(0,+\infty).$
 \label{lem03}
\end{lemma}

\begin{proof}
Note that $f^\#(t,x,\xi)$ is absolutely continuous with respect to
the time $t$ and that the solution $f$ satisfies
\begin{eqnarray}
\frac{\partial}{\partial t}f^\#(t,x,\xi)=
-\Gamma^\#(t,x,\xi)f^\#(t,x,\xi) \hspace*{3.5cm}\nonumber \\
+\eta^\#(t, x, \xi)\int_{{\bf R}^3}\Gamma(t, x+t\xi,
\xi_1)P(-\xi\cdot\xi_1)f(t, x+t\xi, \xi_1)d\xi_1 .
 \label{(3.1.1)}
\end{eqnarray}
Integrating equation (\ref{(3.1.1)}) over the two variables
$(x,\xi)$ and using assumptions (\ref{(2.1.2)}) and (\ref{con04}),
we can know that (\ref{ineqfun01}) holds true. Furthermore, with the
help of (\ref{(2.1.4)}), integrating (\ref{ineqfun01}) over the time
$t$ from zero to $s$ gives (\ref{ineqfun02}).
\end{proof}

\begin{remark}
The inequality (\ref{ineqfun01}) implies that the total mass in UFM
does not increase as time increases.
\end{remark}

Then, using Lemma~\ref{lem03}, we can show that
\begin{theorem}
Under all the assumptions in Lemma~\ref{lem03}, the solution
$f(t,x,\xi)$ to the Boltzmann-like equation (\ref{(2.1.1)})
converges in $L^1({\bf R}^3\times{\bf R}^3)$ to
$f_{\infty}(t,x,\xi)$ as $t\to+\infty,$ where
\begin{eqnarray}
f_{\infty}(t,x,\xi)=f_0(x-t\xi,\xi)-\int_0^{+\infty}\Gamma(\sigma,x+(\sigma-t)\xi,\xi)f(\sigma,x+(\sigma-t)\xi,\xi)d\sigma
\hspace*{0.5cm}\nonumber \\
+\int_0^{+\infty}\int_{{\bf
R}^3}\eta(\sigma,x+(\sigma-t)\xi,\xi)P(-\xi\xi_1)\Gamma(\sigma,x+(\sigma-t)\xi,\xi_1)f(\sigma,x+(\sigma-t)\xi,\xi_1)d\sigma
d\xi_1~~~~~~ \label{(3.1.2)}
\end{eqnarray}
for any point $(t,x,\xi).$
 \label{th03}
\end{theorem}

\begin{proof}
Note that $f$ is a nonnegative solution to the Boltzmann-like
equation (\ref{(2.1.1)}). Then, by (\ref{(3.1.2)}), we can easily
know that
\begin{eqnarray}
\iint_{{\bf R}^3\times{\bf
R}^3}|f(t,x,\xi)-f_{\infty}(t,x,\xi)|dxd\xi\leq\int_t^{+\infty}
\iint_{{\bf R}^3\times{\bf
R}^3}\Gamma(\sigma,x,\xi)f(\sigma,x,\xi)dxd\xi d\sigma
\hspace*{0.5cm}\nonumber \\
+\int_t^{+\infty}\iint_{{\bf R}^3\times{\bf R}^3}\left[\int_{{\bf
R}^3}\eta(\sigma,x,\xi)P(-\xi\xi_1)d\xi_1\right]\Gamma(\sigma,x,\xi_1)f(\sigma,x,\xi_1)dxd\xi
 d\sigma . \label{(3.1.3)}
\end{eqnarray}
By (\ref{con04}), it then follows that
\begin{eqnarray}
\iint_{{\bf R}^3\times{\bf
R}^3}|f(t,x,\xi)-f_{\infty}(t,x,\xi)|dxd\xi\leq
(1+\delta)\int_t^{+\infty}
\iint_{{\bf R}^3\times{\bf
R}^3}\Gamma(\sigma,x,\xi)f(\sigma,x,\xi)dxd\xi d\sigma.
\label{(3.1.4)}
\end{eqnarray}
Since the inequality (\ref{ineqfun02}) in Lemma~\ref{lem03} implies
that the integral on the right side of (\ref{(3.1.4)}) is convergent
to zero as $t\to +\infty,$ the left side of (\ref{(3.1.4)})
converges to zero as time goes to infinity. This completes our proof
of this theorem.
\end{proof}

\begin{remark}
Since $f_{\infty}(t,x,\xi)$ defined by (\ref{(3.1.2)}) describes  a
free motion in UFM, Theorem~\ref{th03} shows that the solution to
the Cauchy problem on UFM is time-asymptotically convergent to the
free motion in $L^1$-norm under some assumptions. It is worth
mentioning that this time-asymptotic convergence of solution still
holds true for the Boltzmann or Enskog equation under some suitable
assumptions (e.g., \cite{p89}).
\end{remark}

\vspace*{0.5cm}\noindent {\bf Acknowledgement.} The authors would
like to thank the referees of this paper for their valuable comments
and suggestions.

\end{document}